\RequirePackage{fix-cm}
\documentclass[smallextended]{svjour3}       
\smartqed  
\usepackage{graphicx}

\usepackage{t1enc}
\usepackage{graphicx}
\usepackage{epstopdf}
\usepackage{amsmath} 
\usepackage{amsbsy} 
\usepackage{cite}
\usepackage{amssymb}
\usepackage{float}
\usepackage{commath}
\usepackage[utf8]{inputenc} 
\usepackage{microtype}
\usepackage{mathptmx}      
\usepackage{latexsym}
 \journalname{Wireless Personal Communications}
\begin{document}

\title{Block Diagonalization Type Precoding Algorithms for IEEE 802.11ac Systems
}


\author{Ishhanie Majumdar  
}


\institute{Ishhanie Majumdar \at
              Indian Institute of technology Madras \\
              Chennai-600 036\\
              \email{ee12s058@ee.iitm.ac.in}           
}


\maketitle

\begin{abstract}
Block diagonalization (BD) based precoding schemes are well-known linear transmit strategies employed in the downlink of multi-user multiple-input multiple-output (MU-MIMO) systems. BD type precoding algorithms employed at the transmit side effect the suppression of multi-user interference (MUI) by the decomposition of MU-MIMO broadcast channel into multiple single-user MIMO (SU-MIMO) channels followed by parallelization of the SU-MIMO channels to obtain independent streams of each user. Given that the design of linear precoding algorithms has made significant progress, the implementation of these techniques in standards for wireless local area networks (WLAN) remains an open question. In this work, schemes for implementation of BD based precoding techniques in the framework of IEEE 802.11ac standard are proposed followed by performance evaluation of these techniques in the proposed framework. I analyze the sum-rate and the bit-error-rate (BER) performance of the techniques in my framework and obtain computational complexity-wise as well as performance-wise optimal algorithm for my system.

\end{abstract}

\section{Introduction}
\label{intro}
Multiple-input multiple-output (MIMO) systems have drawn considerable research interest on account of achieving high channel capacity \cite{MIMOcomm, 1268373}. Research in this area has given particular attention to two optimization problems: throughput maximization and power control \cite{987680,956970,956583}. The capacity in a multiuser network is achieved by maximization of the sum of the information rates for all users subject to a sum power constraint. The problem of power control is based on minimization of the total transmitted power while maintaining a minimum Quality-of-Service (QoS) level for each user in the network. In either case, a satisfactory solution aims to balance high throughput or good QoS at one node in the network at the cost of interference produced at other nodes \cite{1}. Therefore, mitigation of multi-user interference (MUI) is a particular challenge in multiuser systems. Various precoding schemes have been proposed to mitigate MUI in multi-user systems. These schemes exploit the channel state information at the transmitter (CSIT) which allows for joint processing of all users' signals towards performance improvement. The information theoritic results in \cite{997089} have shown that the sum capacity of a MU-MIMO downlink system is reached through dirty paper coding (DPC) or Tomlinson-Harashima precoding (THP). However, these techniques are hard to implement in practice owing to the use of a complex sphere decoder or an approximate closest point solution. Linear precoding provides for a relatively easier implementation in practical systems as opposed to non linear DPC and THP. The generalized methods for linear precoding address various optimization criteria such as minimum mean-squared error (MMSE) and maximum information rate. Moreover, convex optimization based strategies for precoding have drawn efforts in the recent years towards the aim of maximizing system throughput and minimizing power consumption. Convex optimization based precoders for large scale MIMO have been investigated in \cite{wang2014convex} for reducing the peak-to-average power ratio (PAPR) of the signals emitted from base station (BS) antennas.  

High dimensional MIMO configurations have shown dramatic gains in rate and reliability \cite{3}. IEEE 802.11ac \cite{IEEE} considers MIMO configurations with up to 8 transmit and receive antennas. However, the challenge in high-dimensional MIMO systems is the design of suitable precoding algorithms which provide good overall performance and low computational complexity at the same time. The linear precoding techniques: block diagonalization (BD) proposed in \cite{1, 4} and regularized block diagonalization (RBD) proposed in \cite{2} are known to impose a high computational cost. The computational complexity is determined by the number of users and the dimensions of each user's channel matrix \cite{5} in a MU-MIMO scenario. The two algorithms employ two successive singular value decompositions (SVD) iteratively for each user which results in a high computational cost. Extra control overhead is imposed by the two algorithms due to the computation of decoding matrices from the second SVD operation \cite{6}.  
 

Recent work has addressed the question of implementation of the BD-type precoding algorithms with low computational complexity. The generalized zero-forcing (ZF) channel inversion (GZI) method has been proposed in \cite{7} to provide low complexity implementation of the BD precoding. The generalized MMSE channel inversion (GMI) method has also been developed in \cite{7} for the RBD precoding. The first SVD operation in the original BD and the RBD precoding is replaced by less complex QR decomposition in GZI and GMI respectively. A low complexity lattice reduction-aided RBD (LC-RBD-LR) has been proposed in \cite{8, 9} to improve the complexity of the original RBD. A high performance simplified GMI (S-GMI) precoding scheme has been proposed in \cite{5} and developed to achieve low complexity through a common channel inversion for all users. A lattice reduction-aided simplified GMI (LR-S-GMI) algorithm has also been developed in \cite{5} which employs lattice reduction technique to transform the SU-MIMO channels obtained from the S-GMI algorithm to obtain parallelized streams for each user. QR-based BD and Jacket matrix methods have been proposed in \cite{10} which employ QR decomposition and element-wise inverse Jacket matrices respectively towards channel matrix decomposition. A precoding scheme has been proposed in \cite{11} in which the first SVD in BD and RBD is replaced by LQ decomposition through Givens or Householder transformation. Wang et al. in \cite{11} have proved that their proposition outperforms BD and competes RBD in performance and computational complexity. However, the implementation of all these schemes in practice remains an open question. 

 The main contributions of the paper are summarized below:
  \renewcommand\labelenumi{\theenumi)}
 \begin{enumerate}
 \item\noindent The BD-type precoding algorithms are implemented in a orthogonal frequency division multiplexing (OFDM) based framework complying with IEEE 802.11ac standard for wireless local area networks (WLAN).
 
 \item\noindent The algorithms are systematically analyzed and summarized. I review the computational complexity of the algorithms which serves an important criterion for determining the optimal algorithm for the proposed framework.
 
 \item\noindent A simulation study of the algorithms in the given framework, is conducted in terms of bit-error-rate (BER) and achievable sum-rate. A comprehensive analysis is developed for the BER and the sum-rate performances. I determine the BD-type precoding algorithm which shows the optimal performance and computational complexity.  
 \end{enumerate}

 This paper is organized as follows. The system model is described in Section $\mathrm{II}$. A brief discussion of the precoding algorithms considered in this paper is presented in Section $\mathrm{III}$. The system framework is discussed, the simulation results are displayed and performance analysis is developed in Section $\mathrm{IV}$. Conclusions are drawn in Section $\mathrm{V}$.
 
 \emph{Notation}: Vectors and matrices are denoted by lower and upper boldface letters respectively. The transpose, Hermitian transpose and inverse of a matrix $\textbf{\textit{A}}$ are denoted as $\textbf{\textit{A}}^{T}$, $\textbf{\textit{A}}^{H}$ and $\textbf{\textit{A}}^{-1}$ respectively. 
\section*{System Model}
I consider a MU-MIMO downlink channel with a BS equipped with $N_{T}$ transmit antennas and \textit{K} users equipped with $N_{k}$ receive antennas each. The total number of receive antennas in the system is $N_{R} = \sum_{k=1}^{\textit{K}} N_{k}$. The transmit vector for the user \textit{k} is defined as $x_{k} \in \mathbb{C}^{d_{k}\times1}$. Each of the $x_{k}$ is stacked into $\textbf{x} = [\textbf{x}_{1}^T,\textbf{x}_{2}^T,\cdots ,\textbf{x}_{\textit{K}}^T]^{T} \in \mathbb{C}^{d \times1}$ where $d = \sum_{k=1}^{K} d_{k}$. The precoding matrix $\textit{\textbf{P}}_{k} \in \mathbb{C}^{N_{T}\times d_{k}}$ for the user \textit{k} is determined at the transmit side using matrices $\textbf{\textit{P}}^{a}_{k}\in\mathbb{C}^{N_{T}\times d_{k}}$ and $\textbf{\textit{P}}^{b}_{k}\in\mathbb{C}^{d_{k}\times d_{k}}$ such that
\begin{equation}
\textit{\textbf{P}}_{k}=\textbf{\textit{P}}^{a}_{k}\textbf{\textit{P}}^{b}_{k},
\end{equation}
Let me define $\textit{\textbf{P}}^{a}$ and $\textit{\textbf{P}}^{b}$ as
\begin{equation}
\textbf{\textit{P}}^{a}=\left [ \textbf{\textit{P}}^{a}_{1}, \textbf{\textit{P}}^{a}_{2}, \cdots, \textbf{\textit{P}}^{a}_{K} \right ] \in\mathbb{C}^{N_{T}\times d},
\end{equation}
\begin{equation}
{\textbf{\textit{P}}}^{b}=\begin{bmatrix}
{\textbf{\textit{P}}}^{b}_{1} &0  &\cdots  &0 \\ 
 0&{\textbf{\textit{P}}}^{b}_{2}  &\cdots  &0 \\ 
 \vdots &\vdots   &\ddots   &\vdots  \\ 
 0&  0&  0& {\textbf{\textit{P}}}^{b}_{K}
\end{bmatrix} \in\mathbb{C}^{d\times d},
\end{equation}	
The joint precoding matrix \textit{\textbf{P}} given by
\begin{equation}\label{eq:a}
\textit{\textbf{P}} = [ \textit{\textbf{P}}_{1},\hspace{2mm}  \textit{\textbf{P}}_{2},\hspace{2mm} \cdots,\hspace{2mm}  \textit{\textbf{P}}_{\textit{K}}] \in \mathbb{C}^{N_{T}\times d},
\end{equation}
can be written in $\textit{\textbf{P}}^{a}$ and $\textit{\textbf{P}}^{b}$ as 
\begin{equation}
\textbf{\textit{P}}={\textbf{\textit{P}}}^{a}{\textbf{\textit{P}}}^{b}.
\end{equation}
I assume frequency selective channels on account of which we use OFDM for transmission where the same MIMO processing is performed on each subcarrier. The channel matrix of the user \textit{k} for a given subcarrier is defined as $\textit{\textbf{H}}_{k} \in \mathbb{C}^{N_{k}\times N_{T}}$ and the joint channel matrix \textit{\textbf{H}} is expressed as
\begin{equation}\label{eq:b}
\textit{\textbf{H}} = [\textit{ \textbf{H}}_{1}^{T},\hspace{2mm} \textit{\textbf{H}}_{2}^{T},\hspace{2mm}\cdots,\hspace{2mm} \textit{\textbf{H}}_{\textit{K}}^{T}]^{T} \in \mathbb{C}^{N_{R}\times N_{T}} ,
\end{equation}
Each user employs its decoding matrix. The decoding matrix of the user \textit{k} is represented as $\textit{\textbf{D}}_{k} \in \mathbb{C}^{d_{k}\times N_{k}}$ and the joint decoding matrix {\textbf{\textit{D}}} is denoted as
 \begin{equation}
 {\textbf{\textit{D}}}=\begin{bmatrix}
 {\textbf{\textit{D}}}_{1} &0  &\cdots  &0 \\ 
  0&{\textbf{\textit{D}}}_{2}  &\cdots  &0 \\ 
  \vdots &\vdots   &\ddots   &\vdots  \\ 
  0&  0&  0& {\textbf{\textit{D}}}_{K}
 \end{bmatrix} \in\mathbb{C}^{d\times N_{R}},
 \end{equation}
Given the receive vector of the user \textit{k} is denoted by $\textbf{\textit{y}}_{k}\in\mathbb{C}^{d_{k}\times 1}$, I construct the joint receive vector $\textbf{\textit{y}}=[\textbf{\textit{y}}_{1}^{T},\textbf{\textit{y}}_{2}^{T},\cdots,\textbf{\textit{y}}_{\textit{K}}^{T}]^{T}\in\mathbb{C}^{d\times 1}$ as
\begin{equation}\label{eq:c}
\textbf{\textit{y}} = \textbf{\textit{D}}(\textbf{\textit{HPx}} + \textbf{\textit{n}}) .
\end{equation}
where $\textbf{\textit{n}}\in\mathbb{C}^{N_{R}\times 1}$ denotes the additive white Gaussian noise (AWGN) vector. The channel matrix \textbf{\textit{H}} is transformed to a block diagonal matrix when pre-processed and post-processed by \textbf{\textit{P}} and \textbf{\textit{D}} respectively which implies the MU-MIMO downlink channel is decomposed to single-input single-output (SISO) channels and MUI at the users is mitigated.

\section*{Brief Discussion of the Precoding Algorithms}
I begin with the discussion of the fundamental BD algorithm which provides the matrices \textit{\textbf{P}} and \textit{\textbf{D}} under the assumption of no power loading. The BD algorithm computes the precoding matrix in two stages. The first stage of the precoding filter aims at transforming a MU-MIMO channel into SU-MIMO channels such that for each of the users, the interference from other users is suppressed. The second stage aims to decouple the SU-MIMO channels into parallelized streams of each user.

The channel matrix $\overline{\textbf{\textit{H}}}_{k}$ for the user \textit{k} is defined as
\begin {equation}\label{eq:late}
\overline{\textbf{\textit{H}}}_{k} =  [ {\textbf{\textit{H}}_{1}^{T}} \cdots \hspace{1mm}\textbf{\textit{H}}_{k-1}^{T} \hspace{1mm} \textbf{\textit{H}}_{k+1}^T \cdots \hspace{1mm} \textbf{\textit{H}}_{K}^T ]^{T} \in \mathbb{C}^{\overline{N}_{k}\times N_{T}} ,
\end{equation}
where $\overline{N}_{k} = N_{R}-N_{k}$.
The SVD of $\overline{\textbf{\textit{H}}}_{k}$ is given as 
\begin{equation}
\overline{\textbf{\textit{H}}}_{k} =\overline {\textbf{\textit{U}}}_{k} \overline{\boldsymbol{\Sigma}}_{k} \overline{\textbf{\textit{V}}}_{k}^{H} = \overline{\textit{\textbf{U}}}_{k} \overline{\boldsymbol{\Sigma}}_{k} [ \overline{\textit{\textbf{V}}}_{k}^{1} \hspace{2mm} \overline{\textit{\textbf{V}}}_{k}^{0} ]^{H} ,
\end{equation}
where $\overline{\textit{\textbf{U}}}_{k}\in\mathbb{C}^{\overline{N}_{k}\times \overline{N}_{k}} $ and $\overline{\textbf{\textit{V}}}_{k}\in\mathbb{C}^{N_{T}\times N_{T}}$ are full-rank unitary matrices and $\overline{\boldsymbol{\Sigma}}_{k} \in \mathbb{C}^{\overline{N}_{k}\times N_{T}}$ is a diagonal matrix. $\overline{\textbf{\textit{V}}}_{k}^{1}$ and $\overline{\textbf{\textit{V}}}_{k}^{0}$ contain non-zero and zero singular vectors of $\overline{\textbf{\textit{H}}}_{k}$ respectively.

The first precoding filter $\textbf{\textit{P}}_{k}^{a}$ for the user \textit{k} is obtained as a solution to the BD constraint explained in \cite{5, 1468466}. The constraint is imposed such that $\textbf{\textit{P}}_{k}^{a}$ is the basis for the null space of $\overline{\textbf{\textit{H}}}_{k}$ subject to the average transmit power for each user. Therefore,
\begin{equation}\label{eq:k}
\textbf{\textit{P}}_{k}^{a}=\overline{\textbf{\textit{V}}}_{k}^{0} .
\end{equation}
The first step of precoding decomposes the MU-MIMO channel into \textit{K} orthogonal SU-MIMO channels. The effective channel of the user \textit{k} is given by
\begin{equation}
\textbf{\textit{H}}_{\text{eff},k}=\textbf{\textit{H}}_{k}\textbf{\textit{P}}_{k}^{a} ,
\end{equation}
The SVD of the user \textit{k}'s effective channel is given as
\begin{equation}
\textbf{\textit{H}}_{\text{eff},k}=\textbf{\textit{U}}_{k} \boldsymbol{\Sigma}_{k} \textbf{\textit{V}}_{k}^{H}=\textbf{\textit{U}}_{k} \boldsymbol{\Sigma}_{k} [ \textbf{\textit{V}}_{k}^{1} \hspace{2mm} \textbf{\textit{V}}_{k}^{0} ] ,
\end{equation}
If the rank of $\textbf{\textit{H}}_{\text{eff},k}$ is L then $\textbf{\textit{V}}_{k}^{1}$ contains the first L columns of $\textbf{\textit{V}}_{k}$. The second precoding filter is obtained thus,
\begin{equation}\label{eq:l}
\textbf{\textit{P}}_{k}^{b}=\textbf{\textit{V}}_{k}^{1} .
\end{equation}
The precoding matrix $\textbf{\textit{P}}_{k}$ for user \textit{k} is given by
\begin{equation}
\textbf{\textit{P}}_{k}=\textbf{\textit{P}}_{k}^{a}\textbf{\textit{P}}_{k}^{b} .
\end{equation}
The decoding matrix for user \textit{k} is given by
\begin{equation}
\textbf{\textit{D}}_{k}=\textbf{\textit{U}}_{k}^{H} .
\end{equation}

The RBD aims to balance noise with MUI through a regularization factor. It, however, gives rise to residual interferences between the SU-MIMO channels. The RBD algorithm is same as the BD except that its first and second precoding filters are calculated as (\ref{eq:m}) and (\ref{eq:n}).
\begin{equation}\label{eq:m} 
\textit{\textbf{P}}_{k}^{a}= \overline{\textit{\textbf{V}}}_{k}( \overline{\boldsymbol{\Sigma}}_{k}^{T}\overline{\boldsymbol{\Sigma}}_{k} + \alpha \textit{\textbf{I}}_{N_{T}} )^{-1/2} .
\end{equation}
where $\alpha=\sqrt{\frac{N_{R}\sigma^{2}_{n}}{\xi}}$ is the regularization factor and $\sigma^{2}_{n}$ denotes the noise variance and $\xi$ denotes the whole average transmit power.
\begin{equation}\label{eq:n}
\textbf{\textit{P}}_{k}^{b}=\textbf{\textit{V}}_{k} .
\end{equation}

 The GZI obtains its first precoding filter by the zero-forcing channel inversion (ZF-CI) of the joint channel matrix followed by the QR decomposition of the resultant channels of individual users. The procedure for computing its second precoding filter and the decoding filter is same as the BD's.

The pseudo-inverse of the MU-MIMO channel matrix \textbf{\textit{H}} is defined as
\begin{equation}
\hat{\textbf{\textit{H}}}=\textbf{\textit{H}}^{\textbf{\textit{H}}}(\textbf{\textit{H}}\textbf{\textit{H}}^{\textbf{\textit{H}}})^{-1}=[ \hat{\textbf{\textit{H}}}_{1},\hspace{1mm} \hat{\textbf{\textit{H}}}_{2} \cdots \hspace{1mm} \hat{\textbf{\textit{H}}}_{\textit{K}} ] ,
\end{equation}
where user \textit{k}'s matrix $\hat{\textbf{\textit{H}}}_{k} \in \mathbb{C}^{N_{T}\times N_{k}}$.
The QR decomposition of $\hat{\textbf{\textit{H}}}_{k}$ is given by
\begin{equation}
\hat{\textbf{\textit{H}}}_{k} = \hat{\textbf{\textit{Q}}}_{k} \hat{\textbf{\textit{R}}}_{k}, 
\end{equation}
Thus, the first precoding filter for the user \textit{k} is obtained as
\begin{equation}
\textbf{\textit{P}}_{k}^{a} = \hat{\textbf{\textit{Q}}}_{k} .
\end{equation}

Replacing the first SVD of RBD by a less complex QR decomposition reduces the complexity of the RBD. The LC-RBD-LR further replaces the second SVD of the RBD by a complex lattice reduction (CLR) algorithm which has its complexity mainly due to a QR decomposition. In the LC-RBD-LR, CLR is applied to obtain a new SU-MIMO channel basis for each user which is nearly orthogonal as compared to the original matrix. The algorithm does not require to compute a decoding matrix.  

In order to compute $\textbf{\textit{P}}_{k}^{a}$, the channel extension of $\overline{\textbf{\textit{H}}}_{k}$ for user \textit{k} is defined as
\begin{equation}
\overline{\overline{\textbf{\textit{H}}}}_{k}=\left \{ \alpha \textbf{\textit{I}}_{\overline{N}_{k}} , \overline{\textbf{\textit{H}}}_{k} \right \},
\end{equation}
where $\textbf{\textit{I}}_{\overline{N}_{k}}$ is a $\overline{N}_{k} \times \overline{N}_{k}$ identity matrix.
The QR decomposition of $\overline{\overline{\textbf{\textit{H}}}}_{k}^{H}$ is expressed as
\begin{equation}
\overline{\overline{\textbf{\textit{H}}}}_{k}^{H}=\textbf{\textit{Q}}_{k}\textbf{\textit{R}}_{k},
\end{equation}
where $\textbf{\textit{Q}}_{k} \in \mathbb{C}^{\overline{N}_{k} + N_{T} \times \overline{N}_{k} + N_{T}}$ is a unitary matrix and $\textbf{\textit{R}}_{k} \in \mathbb{C}^{\overline{N}_{k} + N_{T} \times \overline{N}_{k}}$ is a upper triangular matrix. Thereby, $\textbf{\textit{P}}_{k}^{a}$ is given by
\begin{equation}
\textbf{\textit{P}}_{k}^{a}=\textbf{\textit{Q}}_{k}\left ( \overline{N}_{k} + 1: \overline{N}_{k} + N_{T} , \overline{N}_{k} + 1: \overline{N}_{k} + N_{T} \right ),
\end{equation}
The CLR transformation is performed on the effective channel matrix of the user \textit{k} after the first level of precoding. The CLR transformation of $\textbf{\textit{H}}_{\text{eff},k}^{T}$ is defined as
\begin{equation}
{\tilde{\textbf{\textit{H}}}}_{\text{eff},k}=\textbf{\textit{U}}_{k}\textbf{\textit{H}}_{\text{eff},k},
\end{equation}
where $\textbf{\textit{U}}_{k}$ is a complex valued unimodular matrix.
The matrix $\tilde{\textbf{\textit{P}}}_{k}^{b}$ is obtained by the ZF-CI of ${\tilde{\textbf{\textit{H}}}}_{\text{eff},k}$ as
\begin{equation}
\tilde{\textbf{\textit{P}}}_{k}^{b}={\tilde{\textbf{\textit{H}}}}^{H}_{\text{eff},k}\left ( {\tilde{\textbf{\textit{H}}}}_{\text{eff},k}{\tilde{\textbf{\textit{H}}}}^{H}_{\text{eff},k} \right )^{-1}.
\end{equation}

The S-GMI employs MMSE channel inversion of the MU-MIMO channel matrix followed by QR decomposition of the SU-MIMO channel matrices of individual users, to obtain its first precoding matrix. The second precoding matrix as well as the decoding matrix are achieved by SVD of the SU-MIMO channels. 

The LR-S-GMI employs the same procedure as the S-GMI's to obtain its first precoding filter, while it obtains its second precoding filter through transformation by lattice reduction (LR) instead of the corresponding SVD operation in the S-GMI. The precoding algorithm does not require decoding matrix, thus the receiver structure is simplified. The Complex Lenstra-Lenstra-Lovasz (CLLL) algorithm proposed in \cite{4787140} is used to implement the LR transformation.

The QR-EVD performs QR decomposition of the channel matrices of each user given by (\ref{eq:late}) to obtain its first precoding filter and eigen value decomposition (EVD) of the SU-MIMO channel matrices to obtain its second precoding filter and the decoding filter in sequel. 

\begin{figure}[!h]
\centering
\includegraphics[width=0.75\textwidth]{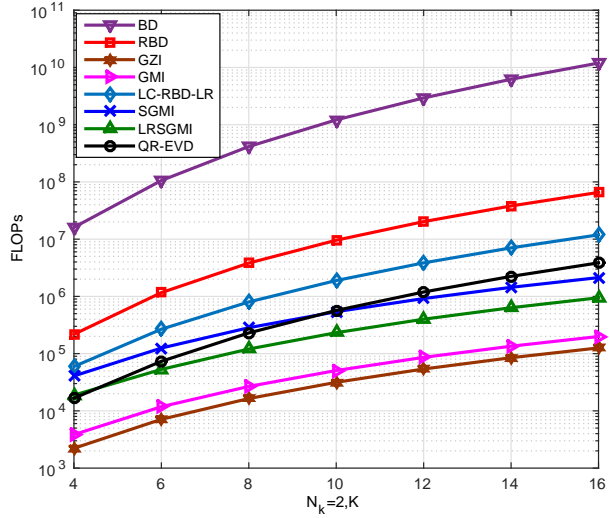}
\caption{Computational complexity in terms of FLOPs for fixed \text{$N_{k}$}.}
\label{fig:FLOPS}
\end{figure}

Fig. \ref{fig:FLOPS} depicts the computational complexity of the precoding algorithms in terms of floating point operations (FLOPs), assuming a fixed receive antenna configuration of $N_{k}=2$ antennas for each user. The conventional BD and the RBD precoding show a faster increase in the computational complexity with the increase of \textit{K} as compared to the other precoding algorithms. This behaviour of the BD and the RBD is because of the first SVD operation which is implemented \textit{K} times on $\overline{\textit{\textbf{H}}}_{k}$ of dimension $\overline{N_{k}} \times N_{T}$. The LC-RBD-LR precoding requires lesser FLOPs though it implements QR decomposition \textit{K} times on $\overline{\textit{\textbf{H}}}_{k}$, because QR decomposition is much simpler than the SVD operation for the same matrix dimensions. The S-GMI shows further reduction in computational complexity as it requires one channel inversion in addition to QR decomposition performed on $\overline{\textit{\textbf{H}}}_{k,mse}$ having a lower dimension $N_{k} \times N_{k}$. The computational complexity of the QR-EVD closely follows that of the S-GMI since the former implements EVD as a substitute of SVD in the latter. The LR-S-GMI requires lesser FLOPs than the S-GMI since it replaces the SVD operation in the latter by a computationally less expensive LR transformation. In this work, I use the knowledge of the computational complexity of the algorithms to determine the optimal algorithm for the framework described in the following section.

\section*{Simulation Results}
I consider an uncoded MU-MIMO system with a BS equipped with $N_{T}=8$ transmit antennas and $\textit{K}=4$ users. Each of the 4 users consists of 2 receive antennas i.e., $N_{k}=2$ where $k = 1,2,\cdots,\textit{K}$ and therefore, $N_{R}=N_{T}$. The BS transmits 2 spatial streams to each of the 4 users, i.e., $d_{k} = 2$. I denote this scenario as $(2,2,2,2)\times 8$ case.
\begin{figure}[!h]
\centering
\includegraphics[width=0.75\textwidth]{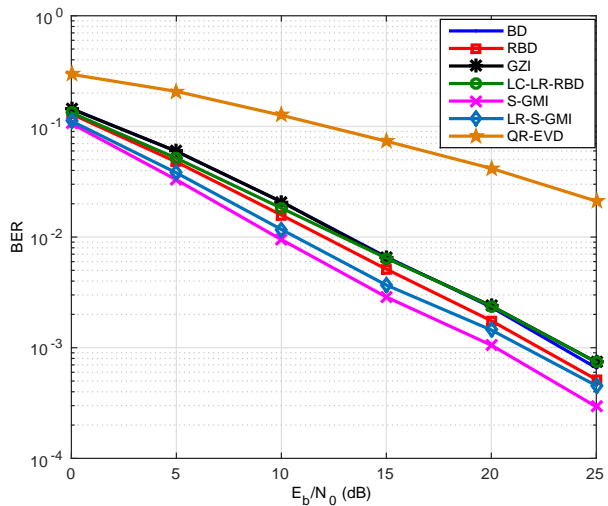}
\centering
\caption{BER performance, (2,2,2,2) x 8 MU-MIMO for 160 MHz 802.11ac channel. }
\label{fig:is}
\end{figure}

I examine the performance of the precoding algorithms in a OFDM based framework \cite{mypaper} compliant to IEEE 802.11ac standard. In the initialization stage of our simulations, I provide the OFDM related parameters and the channel parameters. For an OFDM transmission in a contiguous 160 MHz channel, I consider 512 subcarriers containing 484 data subcarriers, each occupying a bandwidth of 312.5 kHz. I use quadrature phase shift keying (QPSK) modulation for the transmit vector $x_{k}$ intended for the user \textit{k} at a given frequency. The channel matrix \textbf{\textit{H}} of the MU-MIMO environment is modeled as a complex Gaussian with zero mean and unit variance. In order to transform the frequency selective MU-MIMO channel into uncorrelated block fading channels, \textbf{\textit{H}} is obtained for every subcarrier centered around 5.21 GHz as per the standard. The channel impulse response is determined by the delay spread which is calculated based on the indoor range of 35 m. I assume perfect channel estimation at the receive side and error-free feedback channels. Noise variance is obtained using \cite{5} as $\sigma_{n}^{2} = \frac{N_{R}\xi}{N_{T}M(E_{b}/N_{0})}$ where $\xi$ denotes the whole average transmit power and \textit{M} denotes the number of transmitted information bits per channel symbol. My implementation of the precoding schemes assumes no power loading between users and streams for simplicity.

\subsection*{BER Performance}
Fig. \ref{fig:is} compares the BER performance of the precoding algorithms. The BER curve for the GZI closely follows that of the BD which is implied from the fact that both these algorithms completely cancel the MUI and yield comparable signal-to-noise ratios (SNRs). The BER performance of the S-GMI is the best among all. The LR-S-GMI follows the S-GMI closely. The BER performance of the LC-RBD-LR is similar to that of the RBD which can be inferred from the fact that the LC-RBD-LR is designed as a computationally efficient version of the RBD. The QR-EVD precoding has a low performance despite the fact that EVD completely diagonalizes the SU-MIMO channels, since the block diagonalized channel variance is lower compared to that obtained by all the other algorithms.
\begin{figure}[!h]
\centering
\includegraphics[width=0.75\textwidth]{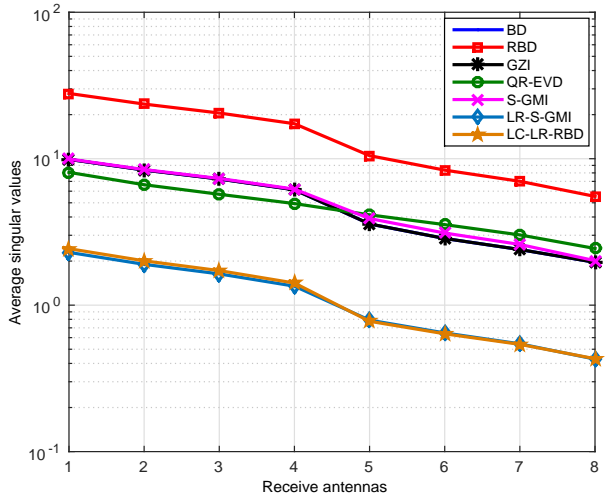}
\caption{Singular values of the block diagonalized channel matrices at SNR=15 dB.}
\label{fig:ss}
\end{figure}

Fig. \ref{fig:ss} provides an explanation for the relative BER behaviour of the precoding algorithms. I consider $E_{b}/N_{0} = 15$ dB for our examination. The BER performance of an algorithm is attributed to the singular values of the block diagonalized channel matrices resulting from that algorithm. The relationship of BER with the singular values of the block diagonalized channel matrices, alternatively the gains of the multiple SISO channels is represented by (\ref{aber}) which shows the BER of QPSK signaling \cite{tse} over a Rayleigh fading channel of gain $\sigma$.
\begin{equation}\label{aber}
p_{e}=\mathrm{Q}\left(\sqrt{\frac{\abs{\sigma}^{2}}{N_{0}}}\right).
\end{equation}
The progressive decay of BER with respect to $\sigma$ as indicated by (\ref{aber}) implies higher the channel gain, lower the BER across the channel. It is therefore implied that a lower BER is attributed to higher singular values of the block diagonalized channel matrices. In the plot of the average singular values of the block diagonalized channel matrices against the channel dimension, a higher position of a graph of an algorithm denotes better performance of the algorithm compared to others.       
 \begin{figure}[!h]
 \centering
 \includegraphics[width=0.75\textwidth]{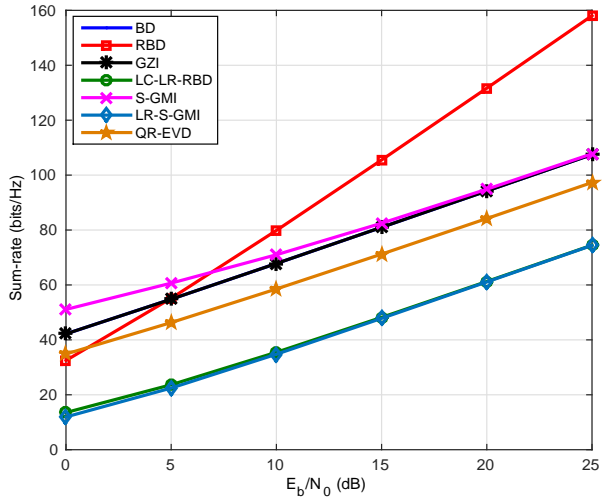}
 \centering
 \caption{Sum-rate performance, (2,2,2,2) x 8 MU-MIMO for 160 MHz 802.11ac channel. }
 \label{fig:maj}
 \end{figure}
 However, there is a slight shift in the order of the graphs of the algorithms in Fig. \ref{fig:ss} compared to Fig. \ref{fig:is} which is because the OFDM parameters are not considered in Fig. \ref{fig:ss} while they are considered in Fig. \ref{fig:is}. Throughout my simulations I consider 100 MIMO channel realizations. Singular values are averaged over all the subcarriers and channel realizations to obtain Fig. \ref{fig:ss}.
\subsection*{Sum-rate Performance}
Fig. \ref{fig:maj} illustrates the sum-rate performance of the precoding algorithms. The calculation of sum-rate \cite{5} is given by 
\begin{equation}\label{eq:gff}
C = \log\left(\det\left(\textbf{\textit{I}} + \sigma^{-2}_{n} \textit{\textbf{H}} \textit{\textbf{P}} \textit{\textbf{P}}^{H} \textit{\textbf{H}}^{H}\right)\right) \hspace{2.5mm} \text{(bits/Hz)}. 
\end{equation}
It can be observed that the maximum achievable sum-rate of the LR-S-GMI is almost the same as the LC-RBD-LR's. This behaviour can be attributed to the fact that both the algorithms employ lattice reduction which has a major impact on the overall precoding matrix in these algorithms. The RBD achieves the best sum-rate performance at high SNRs. The S-GMI shows a small but significant improvement over the concurrent performances of the BD and the GZI at low SNRs and a marginal improvement at high SNRs.
\begin{figure}[!h]
\centering
\includegraphics[width=0.75\textwidth,trim={0 0 6cm 18cm},clip]{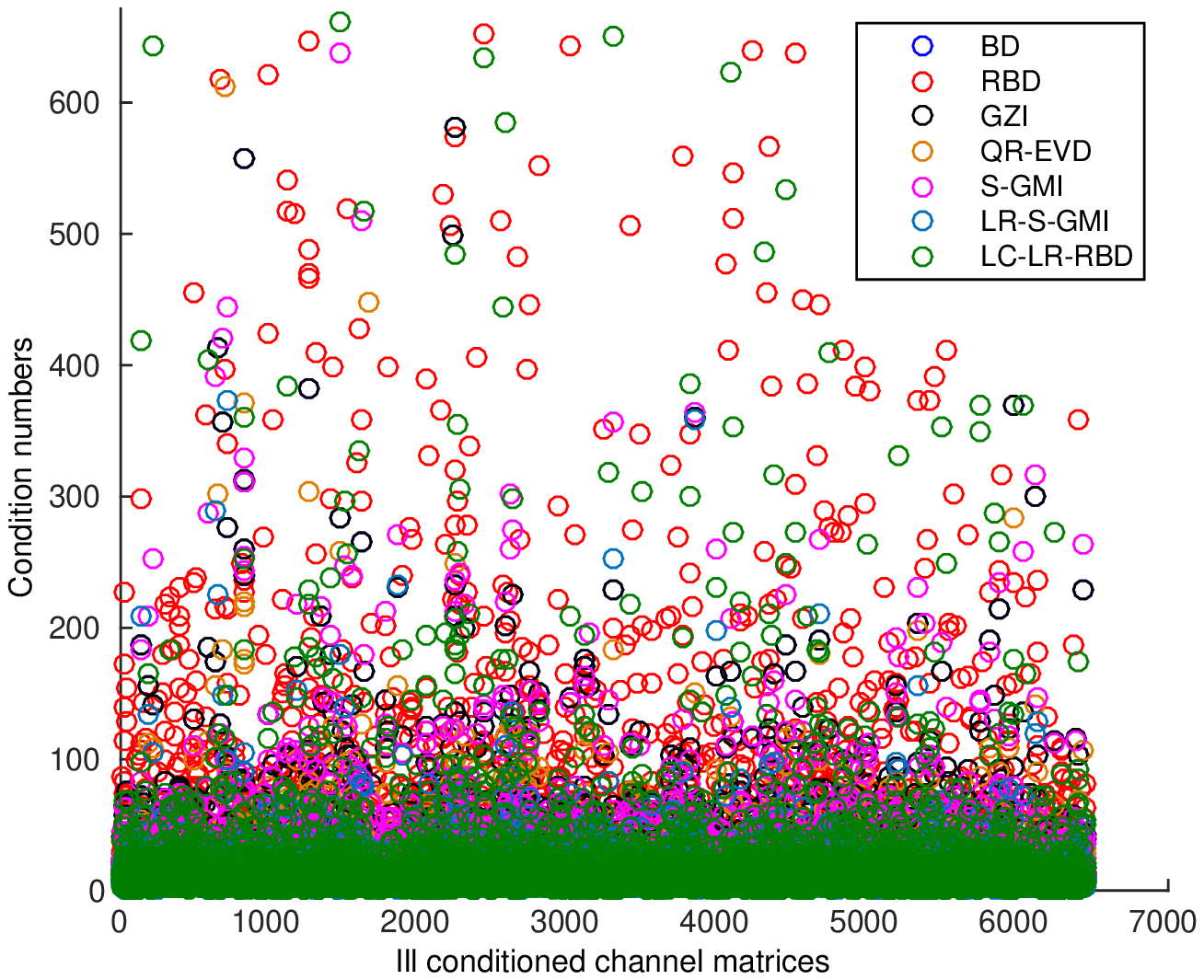}
\centering
\caption{Scatter plot of condition numbers of block diagonalized ill conditioned channels matrices at SNR=15 dB.}
\label{fig:ishaa}
\end{figure} 

I examine the sum-rate performance of the algorithms in the high SNR regime by the eigenspread of the block diagonalized channel matrices resulting from these algorithms. The relationship of sum-rate, otherwise, capacity, with the eigenspread of a block diagonalized channel matrix is explained by (\ref{bj}) and (\ref{cj}). I rewrite (\ref{eq:gff}) as
\begin{equation}\label{bj}
C=\sum_{k=1}^{N_T}\log\left(1+\frac{P_{T} \sigma^2_k}{N_T \sigma^{2}_{n}}\right),
\end{equation}
where the transmit power per antenna is normalized by considering the total transmit power equal to the total number of transmit antennas, i.e., $P_{T}=N_{T}$ and the singular values of the overall channel matrix are denoted by $\sigma_k$. It is to be noted that the normalization of transmit power per antenna is considered throughout my simulations. By Jensen's inequality \cite{tse},
\begin{equation}\label{cj}
\frac{1}{N_T}\sum_{k=1}^{N_T}\log\left(1+\frac{P_{T} \sigma^2_k}{N_T \sigma^{2}_{n}}\right)\leq\log\left(1+\frac{P_{T}}{N_T \sigma^{2}_{n}}\left(\frac{1}{N_T}\sum_{k=1}^{N_T}\sigma^2_k\right)\right). 
\end{equation}
where $\sum_{k=1}^{N_T}\sigma^2_k$ is the total power gain of the channel. It is evident from (\ref{cj}) that the maximum capacity is achieved when all singular values are equal. In general, lower spread of singular values implies larger capacity in high SNR regime. It is therefore implied that lower the eigenspread of a block diagonalized channel matrix, better the sum-rate performance of the algorithm. The better performance of the S-GMI than the BD and the GZI as depicted in Fig. \ref{fig:maj}, is due to the relatively lower eigenspread of the resultant channel matrices.
\begin{figure}[!h]
\centering
\includegraphics[width=0.75\textwidth]{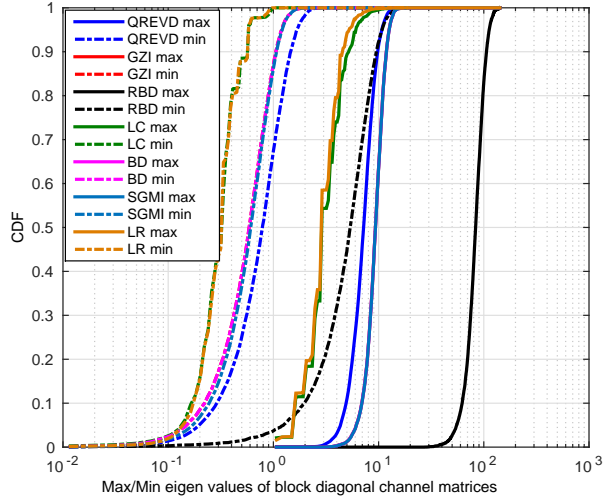}
\caption{CDF plot of maximum and minimum eigen values of block diagonalized ill conditioned channel matrices at SNR=15 dB.}
\label{fig:ishaaa}
\end{figure}

However, I take particular interest in the sum-rate performance of the RBD which shows an intriguing behaviour. I attempt to explain this behaviour of RBD relative to the other algorithms through the condition numbers of the block diagonalized channel matrices resulting from these algorithms. In order to distinguish well conditioned matrices from the ill conditioned, I assume a condition number threshold of 10. The ill conditioned channel matrices in my simulations have condition numbers greater than 10. Fig. \ref{fig:ishaa} shows a scatter plot of condition numbers against the number of ill conditioned channel matrices resulting from each of the precoding algorithms at $E_{b}/N_{0} = 15$ dB. I consider overall channel matrices for each of the subcarriers and channel realizations. Since eigenspread becomes crucial for ill conditioned channel matrices, I consider these matrices for my analysis. It can be seen from Fig. \ref{fig:ishaa} that the condition numbers are concentrated within 200. Therefore, the number of ill conditioned channel matrices having condition number less than or equal to 200 are calculated corresponding to each of the seven precoding algorithms. It is found that the RBD has the least number of such ill conditioned channel matrices which provides a justification of its distinct behaviour relative to the other precoding algorithms.     

I consider an alternate representation of the eigenvalues of the block diagonalized channel matrices in order to establish the sum-rate performance of the precoding algorithms. Fig. \ref{fig:ishaaa} provides an explanation for the behaviour of the algorithms depicted in Fig. \ref{fig:maj}. Fig. \ref{fig:ishaaa} shows a cumulative distribution function (CDF) plot of the maximum and minimum eigenvalues of the block diagonalized channel matrices for each of the algorithms at $E_{b}/N_{0} = 15$ dB. The maximum and minimum eigen values are considered for the ill conditioned channel matrices. The eigenspread is measured as the width between the CDF of the maximum eigenvalue and that of the minimum eigenvalue at the CDF value of 0.5 which is the coordinate of symmetry. It is calculated that this width is the least for RBD by and large. S-GMI follows RBD in the ascending order of the width.

\section*{Conclusion}
In this paper, I have investigated block diagonalization type precoding algorithms for MU-MIMO systems wherein my major contribution has been the implementation of these precoding algorithms in a unified OFDM based framework complying with the IEEE 802.11ac specifications. I have presented the performance comparison and analysis of the precoding algorithms in the proposed framework. Taking into account the BER, the sum-rate performance and the computational complexity of these algorithms, I conclude that the optimal performance is achieved by S-GMI for my system. My current implementation does not consider the problem of power loading. Efficient power allocation schemes can be investigated and applied to the precoding techniques considered in this work. Further extension to this work can be carried out by applying convex optimization strategies to the problems of precoding keeping this system under consideration.



\begin{acknowledgements}
The author thanks Prof. Harishankar Ramachandran for useful discussions on this work.
\\

\noindent \textbf{Conflict of Interest}: The author declares that she has no conflict of interest.

\end{acknowledgements}

\bibliographystyle{ieeetran}
\bibliography{ishbmc} 

\begin{thebibliography}{10}
\providecommand{\url}[1]{#1}
\csname url@samestyle\endcsname
\providecommand{\newblock}{\relax}
\providecommand{\bibinfo}[2]{#2}
\providecommand{\BIBentrySTDinterwordspacing}{\spaceskip=0pt\relax}
\providecommand{\BIBentryALTinterwordstretchfactor}{4}
\providecommand{\BIBentryALTinterwordspacing}{\spaceskip=\fontdimen2\font plus
\BIBentryALTinterwordstretchfactor\fontdimen3\font minus
  \fontdimen4\font\relax}
\providecommand{\BIBforeignlanguage}[2]{{%
\expandafter\ifx\csname l@#1\endcsname\relax
\typeout{** WARNING: IEEEtran.bst: No hyphenation pattern has been}%
\typeout{** loaded for the language `#1'. Using the pattern for}%
\typeout{** the default language instead.}%
\else
\language=\csname l@#1\endcsname
\fi
#2}}
\providecommand{\BIBdecl}{\relax}
\BIBdecl

\bibitem{MIMOcomm}
D.~G. Arogyaswami~Paulraj, Rohit~Nabar, \emph{Introduction to Space-Time
  Wireless Communications}, 1st~ed.\hskip 1em plus 0.5em minus 0.4em\relax
  Cambridge: Cambridge University Press, 2003.

\bibitem{1268373}
I.~Lee, A.~M. Chan, and C.~E.~W. Sundberg, ``Space-time bit-interleaved coded
  modulation for ofdm systems,'' \emph{IEEE Transactions on Signal Processing},
  vol.~52, no.~3, pp. 820--825, March 2004.

\bibitem{987680}
R.~W. Heath, M.~Airy, and A.~J. Paulraj, ``Multiuser diversity for mimo
  wireless systems with linear receivers,'' in \emph{Conference Record of
  Thirty-Fifth Asilomar Conference on Signals, Systems and Computers
  (Cat.No.01CH37256)}, vol.~2, Nov 2001, pp. 1194--1199 vol.2.

\bibitem{956970}
R.~S. Blum, J.~H. Winters, and N.~R. Sollenberger, ``On the capacity of
  cellular systems with mimo,'' in \emph{IEEE 54th Vehicular Technology
  Conference. VTC Fall 2001. Proceedings (Cat. No.01CH37211)}, vol.~2, 2001,
  pp. 1220--1224 vol.2.

\bibitem{956583}
M.~F. Demirkol and M.~A. Ingram, ``Power-controlled capacity for interfering
  mimo links,'' in \emph{IEEE 54th Vehicular Technology Conference. VTC Fall
  2001. Proceedings (Cat. No.01CH37211)}, vol.~1, Oct 2001, pp. 187--191 vol.1.

\bibitem{1}
Q.~Spencer, A.~Swindlehurst, and M.~Haardt, ``Zero-forcing methods for downlink
  spatial multiplexing in multiuser mimo channels,'' \emph{Signal Processing,
  IEEE Transactions on}, vol.~52, no.~2, pp. 461--471, Feb 2004.

\bibitem{997089}
S.~Vishwanath, N.~Jindal, and A.~Goldsmith, ``On the capacity of multiple input
  multiple output broadcast channels,'' in \emph{Communications, 2002. ICC
  2002. IEEE International Conference on}, vol.~3, 2002, pp. 1444--1450 vol.3.

\bibitem{wang2014convex}
S.~Wang, Y.~Li, and J.~Wang, ``Convex optimization based downlink precoding for
  large-scale mimo,'' in \emph{Wireless Communications and Networking
  Conference (WCNC), 2014 IEEE}.\hskip 1em plus 0.5em minus 0.4em\relax IEEE,
  2014, pp. 218--223.

\bibitem{3}
F.~Rusek, D.~Persson, B.~K. Lau, E.~Larsson, T.~Marzetta, O.~Edfors, and
  F.~Tufvesson, ``Scaling up mimo: Opportunities and challenges with very large
  arrays,'' \emph{Signal Processing Magazine, IEEE}, vol.~30, no.~1, pp.
  40--60, Jan 2013.

\bibitem{IEEE}
IEEE, \emph{Wireless LAN Medium Access Control (MAC) and Physical Layer (PHY)
  Specifications. Enhancements for Very Higher Throughput for Operations in
  Bands below 6GHz, IEEE Std 802.11ac$^{TM}$-2013}, 2013.

\bibitem{4}
L.-U. Choi and R.~Murch, ``A transmit preprocessing technique for multiuser
  mimo systems using a decomposition approach,'' \emph{Wireless Communications,
  IEEE Transactions on}, vol.~3, no.~1, pp. 20--24, Jan 2004.

\bibitem{2}
V.~Stankovic and M.~Haardt, ``Generalized design of multi-user mimo precoding
  matrices,'' \emph{Wireless Communications, IEEE Transactions on}, vol.~7,
  no.~3, pp. 953--961, March 2008.

\bibitem{5}
K.~Zu, R.~de~Lamare, and M.~Haardt, ``Generalized design of low-complexity
  block diagonalization type precoding algorithms for multiuser mimo systems,''
  \emph{Communications, IEEE Transactions on}, vol.~61, no.~10, pp. 4232--4242,
  October 2013.

\bibitem{6}
C.-B. Chae, S.~Shim, and R.~Heath, ``Block diagonalized vector perturbation for
  multiuser mimo systems,'' \emph{Wireless Communications, IEEE Transactions
  on}, vol.~7, no.~11, pp. 4051--4057, November 2008.

\bibitem{7}
H.~Sung, S.-R. Lee, and I.~Lee, ``Generalized channel inversion methods for
  multiuser mimo systems,'' \emph{Communications, IEEE Transactions on},
  vol.~57, no.~11, pp. 3489--3499, Nov 2009.

\bibitem{8}
K.~Zu and R.~de~Lamare, ``Low-complexity lattice reduction-aided regularized
  block diagonalization for mu-mimo systems,'' \emph{Communications Letters,
  IEEE}, vol.~16, no.~6, pp. 925--928, June 2012.

\bibitem{9}
K.~Zu, R.~de~Lamare, and M.~Haardt, ``Lattice reduction-aided regularized block
  diagonalization for multiuser mimo systems,'' in \emph{Wireless
  Communications and Networking Conference (WCNC), 2012 IEEE}, April 2012, pp.
  131--135.

\bibitem{10}
\BIBentryALTinterwordspacing
M.~Khan, K.~Cho, M.~Lee, and J.-G. Chung, ``\BIBforeignlanguage{English}{A
  simple block diagonal precoding for multi-user mimo broadcast channels},''
  \emph{\BIBforeignlanguage{English}{EURASIP Journal on Wireless Communications
  and Networking}}, vol. 2014, no.~1, 2014. [Online]. Available:
  \url{http://dx.doi.org/10.1186/1687-1499-2014-95}
\BIBentrySTDinterwordspacing

\bibitem{11}
H.~Wang, L.~Li, L.~Song, and X.~Gao, ``A linear precoding scheme for downlink
  multiuser mimo precoding systems,'' \emph{Communications Letters, IEEE},
  vol.~15, no.~6, pp. 653--655, June 2011.

\bibitem{1468466}
M.~Joham, W.~Utschick, and J.~A. Nossek, ``Linear transmit processing in mimo
  communications systems,'' \emph{IEEE Transactions on Signal Processing},
  vol.~53, no.~8, pp. 2700--2712, Aug 2005.

\bibitem{4787140}
Y.~H. Gan, C.~Ling, and W.~H. Mow, ``Complex lattice reduction algorithm for
  low-complexity full-diversity mimo detection,'' \emph{Signal Processing, IEEE
  Transactions on}, vol.~57, no.~7, pp. 2701--2710, July 2009.

\bibitem{mypaper}
I.~Majumdar and D.~Jalihal, ``Implementation of block diagonalization type
  precoding algorithms for ieee 802.11ac systems,'' in \emph{2015 Fifth
  International Conference on Advances in Computing and Communications
  (ICACC)}, Sept 2015, pp. 200--203.

\bibitem{tse}
D.~Tse and P.~Viswanath, \emph{Fundamentals of Wireless Communication}.\hskip
  1em plus 0.5em minus 0.4em\relax New York: Cambridge University Press, 2005.

\end{thebibliography}

\end{document}